\documentclass[10pt, conference,letterpaper]{IEEEtran}

\IEEEoverridecommandlockouts
\usepackage[colorlinks,pdfstartview=FitH,citecolor=blue,linkcolor=blue,urlcolor=blue]{hyperref}
\usepackage{cite}
\usepackage{amsmath,amssymb,amsfonts}
\usepackage{algorithmic}
\usepackage{graphicx}
\usepackage{subcaption}
\usepackage{textcomp}
\usepackage{xcolor}
\usepackage{float}
\usepackage[export]{adjustbox}
\usepackage{hyperref}
\usepackage{comment}
\usepackage{wrapfig}
\usepackage{dsfont}

\usepackage[
raggedright
]{sidecap}   
\sidecaptionvpos{figure}{t} 
\def\BibTeX{{\rm B\kern-.05em{\sc i\kern-.025em b}\kern-.08em
    T\kern-.1667em\lower.7ex\hbox{E}\kern-.125emX}}

\usepackage{enumitem}
\usepackage{eso-pic}

\usepackage{bbm}

\newcommand\MyIncludeGraphics[2][]{
    \IfFileExists{#2}{%
        \includegraphics[#1]{#2}%
    }{%
        \missingfigure[figwidth=7.0cm]{Missing #2}%
    }%
}%

\DeclareMathOperator{\concat}{+\kern -0.4em+}

\newcommand{\newVar}[2]{\newcommand{#1}{\ensuremath{#2}\xspace}}
  \newVar{\server}{S}
  \newVar{\client}{C}
  \newVar{\rclient}{R_c}
  \newVar{\rserver}{R_s}

\newcommand{\ignore}[1]{}

\begin{document}
\AddToShipoutPictureBG*{%
  \AtTextLowerLeft{%
    \raisebox{-2\height}{%
      \parbox{\textwidth}{\centering%
      \copyright 2023 IEEE. Personal use of this material is permitted.\\ 
      Permission from IEEE must be obtained for all other uses, in any current or future media, including reprinting/republishing this material for advertising or promotional purposes, creating new collective works, for resale or redistribution to servers or lists, or reuse of any copyrighted component of this work in other works.} 
      \hspace{\columnsep}\makebox[\columnwidth]{}
    }%
  }%
}
\author{
    \IEEEauthorblockN{
    Oleg Kolosov\IEEEauthorrefmark{1},
    Mehmet Fatih Akta\c{s}\IEEEauthorrefmark{3},
    Emina Soljanin\IEEEauthorrefmark{3},
    Gala Yadgar\IEEEauthorrefmark{1}
    }
    \IEEEauthorblockA{
    \IEEEauthorrefmark{1}\textit{\{kolosov,gala\}@cs.technion.ac.il, Computer Science, Technion, Israel}}
    \IEEEauthorblockA{
    \IEEEauthorrefmark{3}\textit{    mfatihaktas@gmail.com, emina.soljanin@rutgers.edu, Electrical and Computer Engineering, Rutgers University, USA}}
}

\title{Theory vs.\ Practice in Modeling\\Edge Storage Systems}
\maketitle
\begin{abstract}
Edge systems promise to bring data and computing closer to the users of time-critical applications. Specifically, \textit{edge storage systems} are emerging as a new system paradigm, where users can retrieve data from small-scale servers inter-operating at the network's edge. The analysis, design, and optimization of such systems require a tractable model that will reflect their costs and bottlenecks. Alas, most existing mathematical models for edge systems focus on stateless tasks, network performance, or isolated nodes and are inapplicable for evaluating edge-based storage performance. 

We analyze the \textit{capacity-region model}---the most promising model proposed so far for edge storage systems. The model addresses the system’s ability to serve a set of user demands. Our analysis reveals five inherent gaps between this model and reality, demonstrating the significant remaining challenges in modeling storage service at the edge. 
\end{abstract}
\section{Introduction}\label{sec:introduction}
Today, we are witnessing the emergence of a new system model driven by the IoT revolution. A surge of applications based on smart \textit{edge devices}, such as smart cities and homes, autonomous vehicles, online video gaming, virtual and augmented reality, and machine learning, generate large amounts of data, continuously collected, aggregated, and analyzed by their back-end service. \textit{Edge services} are envisioned to provide storage and compute infrastructure at inter-operating \textit{edge nodes}: micro-datacenters located one or two network hops, approximately several milliseconds, from the end user~\cite{FOGCOMPUTING:ShiD16,chiang2016fog,FOGCOMPUTING:Byers17,Mahmud2018}, each limited to a typical capacity of 50-150 kW and a diameter of approximately 10 feet~\cite{StateOfTheEdge21}.

In \textit{edge storage services}, edge nodes store data objects and serve user requests for these objects. The edge system's ability to provide the expected user-perceived latency greatly depends, as in other distributed computing systems, on the availability and performance with which it can serve these requests. This performance, in turn, depends on the number, capacity, and location of the edge nodes, the allocation of data objects within the system, and the mechanisms used to route and serve user requests. 

Designing and optimizing systems and their components traditionally relied on mathematical abstractions to model their performance. The systems community has been using and refining mathematical models for, e.g., cache management~\cite{WorkingSet}, hard-drive failures and recovery~\cite{Meaningless,FaultLocalityOptimality,Tiger,Azure}, SSD garbage collection and bit-errors~\cite{AnalyticModels,flash_BER1}, straggler mitigation~\cite{ananthanarayanan2013effective,ananthanarayanan2010reining}, and queuing~\cite{gardner2016s&x}. Data access times were the primary motivation for many of these modeling efforts. 

Modeling data access costs and performance in edge systems is challenging for the following reasons: 1)  Edge computing is tightly coupled with 5G technology, which suffers from a ``go-slow cycle'' of user adoption, deployment, and development~\cite{5GEdge}. Thus, very little data is available on how edge systems will be deployed, managed, and used~\cite{benchmarking}. 2) Storage workloads will likely be characterized by heavy-tail distributions similar to those of their cloud-based counterparts. However, user mobility and geo-locality will further increase the dynamic nature of edge access patterns.
3) Edge nodes can easily become unavailable due to limited computing power and network bandwidth, as well as transient connectivity issues~\cite{puliafito2017fog, hao2017challenges}. Existing mathematical models for the edge focus on aspects related to placement and scheduling of jobs~\cite{ ModelingEvaluationOfThreeTierCloudOfThings:LiSD17, ModelingSimulatingFog:EtemadAS17, ResManForMultiUserEdgeComputing:MaoZS17,  LoadBalancingFogComputing:FanA18} 
or on various aspects of the network's performance~\cite{gardner2016s&x}. These models are not directly applicable to modeling storage workloads and services.


The recently proposed \emph{capacity-region model}~\cite{CapacityRegion} makes a first step in analytical modeling of storage at the edge. This simple model calculates the fundamental bounds of an edge system with data objects and redundancy (in the form of erasure coding or replication). In a nutshell, given a collection of nodes and the objects they store, the model calculates the system's \textit{service-capacity region}---the space of user workloads that the system can serve in terms of its total request rate and distribution of object popularities. 

This work is the first attempt to compare an edge storage-service model to reality. We compare the model’s predictions on whether a system can serve a collection of user workloads to the results of a detailed edge-system simulator and a small-scale experimental system. Our evaluation demonstrates the strengths of the capacity-region model, but also highlights several gaps between the model and the realistic systems. These gaps represent five inherent challenges in modeling edge-based storage: the complexity of multiple inter-dependent queues, access-rate representation granularity, sensitivity to specific component states, the effects of geo-locality, and the effect of scale. The successful use of mathematical models in the design and optimization of large-scale edge systems strongly depends on addressing these challenges. 



\section{The Capacity-Region Model}
\label{sec:model}

\begin{figure}[t]
\centering
{\includegraphics[width=0.4\textwidth]{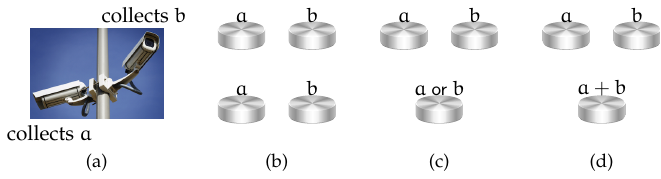}}
{\caption{Edge system with two cameras generating two objects (a) and possible storage models (b-c-d).
\label{fig:storage_eg}}}
\vspace{-1mm}
\end{figure}

\begin{figure}[t]
\vspace{-1mm}
  \centering
 \subfloat[$S(a, a, b, b)$]{
    \centering
\includegraphics[width=0.22\columnwidth,keepaspectratio=true]{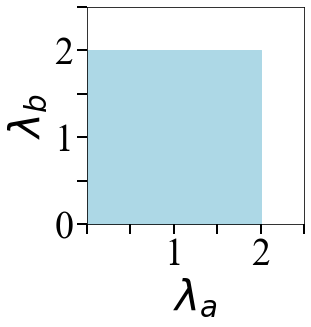}
}\quad
 \subfloat[$S(a, a, b)$]{
    \centering
\includegraphics[width=0.22\columnwidth,keepaspectratio=true]{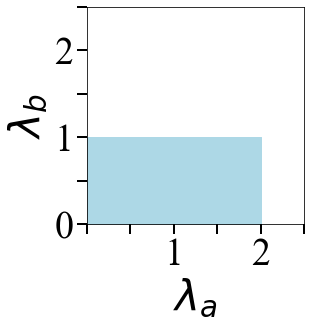}
}\quad
 \subfloat[$S(a, b, a\oplus b)$]{
    \centering
    \includegraphics[width=0.22\columnwidth,keepaspectratio=true]{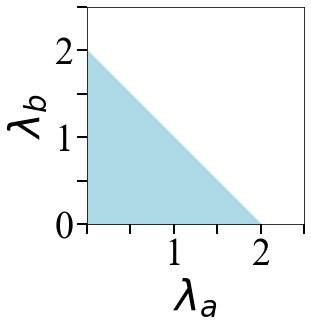}
}
   \vspace{-1mm}
  \caption{The capacity region of the alternative toy systems. 
  }
  \label{fig:fig_capacity_examples}
  \vspace{-3mm}
\end{figure}


We give a brief overview of the capacity-region model, presented and analyzed formally in~\cite{CapacityRegion}. 
The system stores $k$ immutable data objects across $n$ nodes, possibly with redundancy. Each node's \textit{storage capacity} determines the number of objects it can store, and its \textit{service capacity} $\mu$ determines the average rate at which it can serve read requests arriving for its stored objects. For simplicity, we assume here that the objects are of equal size and that all the nodes have the same storage and service capacities. 
Incoming requests at each node are placed into a queue with an unlimited buffer and served in first-in first-out order. The \textit{storage overhead} of the system is the ratio between the cumulative storage capacity of its nodes, and the total size of the $k$ data objects.

The users issue \textsc{read} \emph{requests} for the objects. We refer to the rate of request arrivals for object $i$ as the \emph{request rate} for the object, denoted by $\lambda_i$.
The \emph{demand vector}, or \emph{demand} in short, is a vector $\mathbf{d}$ of $k$ request rates $\mathbf{d} = [\lambda_1, \dots, \lambda_k]$.
The \textit{popularity} of an object $i$ is the fraction $p_i$ of requests arriving for object $i$. We express the demand with $p_i$ as $\mathbf{d} = \lambda [p_1, \dots, p_k]$, where $\lambda$ is the \emph{cumulative demand} $\sum_{i=1}^{k} \lambda_i$.
The model allows for dynamic demands: we view each demand vector as an instance of a collection of possible user behaviors, and the object popularities are not known in advance.

We assume that the storage capacity is sufficient to store all the objects, and some \textit{popular} objects will be requested with a higher probability than others. Thus, the excess storage capacity, if there is any, can be used for storing popular objects redundantly. The model defines two redundancy schemes. The first scheme is \textit{replication}, where multiple copies (replicas) of the object are stored, each on a different server. A request for a \textit{replicated object} can be served by any server storing its replica. The second scheme is \emph{coding}, where redundant objects are created by combining multiple objects. Coded objects can be created with XOR, e.g., 
$a \oplus b$. 
Coded objects can be used to retrieve the original objects from two nodes working \textit{collaboratively}. For example, $a$ can be obtained by downloading $b$ and $a \oplus b$ and then using them to recover $a$. 

Consider the toy system depicted in ~\autoref{fig:storage_eg} with two objects, $a$ and $b$, and edge nodes that can each store one object and handle one request at a time. The users issue three types of simultaneous requests for two objects: $[a,a]$, $[a,b]$, or $[b,b]$. 
With four nodes, the system can store two copies for both objects (denoted as $S(a,a,b,b)$) and handle any request type. With three nodes, the system uses less storage, but can still store two copies of one object. For example, if $a$ is known to be the more popular object, it will be replicated. The resulting system, $S(a,a,b)$, will be able to handle two of the three request types, $[a,a]$ and $[a,b]$. Unfortunately, even though object popularities are often known to be skewed, the identity of the popular objects is not known in advance, and might change continuously. 

Another option is to use three nodes to store the original objects and their \textit{parity}, i.e., $S(a,b,a \oplus b)$, which allows the system to handle any of the three request types. When the requested objects are different ($[a,b]$), each request will be served by the node storing its object.
When the requested objects are the same ($[a,a]$ or $[b,b]$), one request will be served by the node storing the object and the other will be served collaboratively by the other two nodes. The group of objects that can be used together to handle a request is called a \textit{recovery set}. 
The \textit{service cost} of a request is the number of objects downloaded in order to serve it, and the service cost of a demand vector is the cumulative download rate needed to serve it.

According to the model, the system \textit{covers} a demand vector if and only if all user requests are completed within a finite amount of time. 
The system's \textit{capacity region} is defined as the set of all demand vectors that it covers~\cite{CapacityRegion}. \autoref{fig:fig_capacity_examples} illustrates the capacity region of the three toy systems discussed above. 
For example, the system in \autoref{fig:fig_capacity_examples}(b) covers every demand in which $\lambda_a \leq 2\mu$ and $\lambda_b \leq \mu$. Similarly, the system in \autoref{fig:fig_capacity_examples}(c) covers all demand vectors in which $\lambda_a + \lambda_b \leq 2\mu$.

The capacity region is calculated by solving a convex optimization problem. Effectively, the result is similar to assigning requests primarily to the nodes storing the object or its replicas. If all these servers have reached their capacity limit, then the remaining requests for this object are distributed to the servers storing their recovery sets.

\begin{figure}
\centering
\includegraphics[width=0.65\linewidth]{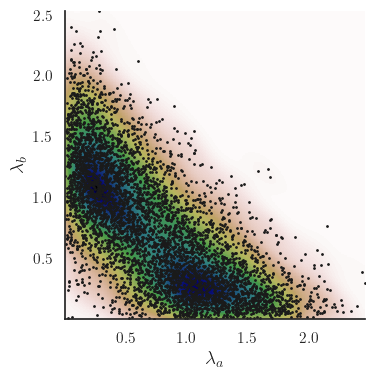}
   \caption{Illustration of 5000 demand vectors generated from a distribution of two objects with Zipf popularity, where $\lambda \sim {\mathcal N}^+(1.5, 0.4)$, and $\alpha \sim {\mathcal N}^+(1, 2)$. 
      \label{fig:fig_eg_pop_heatmap}
}
 \vspace{-0.5cm}
\end{figure}

\begin{figure*}[t!]
  \centering
    \subfloat[$S(a,a,b)$]{
    \centering
\includegraphics[width=0.2\linewidth,keepaspectratio=true]{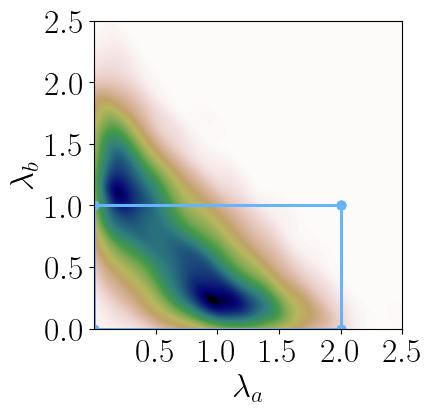}}
\subfloat[$S(a,b,a+b)$]{
    \centering
\includegraphics[width=0.2\linewidth,keepaspectratio=true]{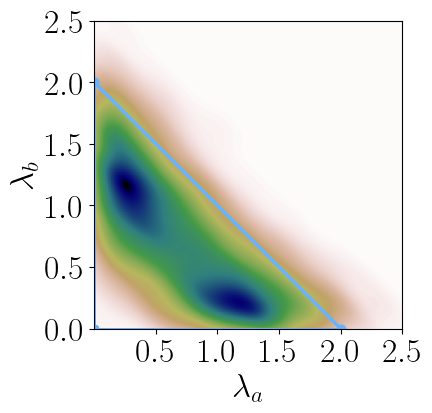}
    }
\subfloat[$S(a,a,b,b)$]{
    \centering
\includegraphics[width=0.2\linewidth,keepaspectratio=true]{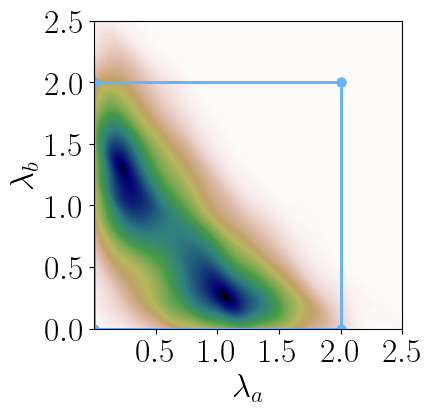}
    }
  \vspace{-1ex}
  \caption{A comparison of the demand distribution and the service region of the systems in \autoref{fig:fig_capacity_examples}. The heatmap is generated using the model in \autoref{fig:fig_eg_pop_heatmap}.
  }
  \label{fig:fig_coveredmass_examples}
  \vspace{-2mm}
\end{figure*}

Recall that in realistic system settings, the precise demand vector is not known in advance. However, the expected user demand can often be characterized by a parametrized distribution. \autoref{fig:fig_eg_pop_heatmap} illustrates the demand vectors generated for a data set of objects $a$ and $b$. These vectors are drawn from a distribution with $\lambda \sim {\mathcal N}^+(1.5, 0.4)$, and Zipf $\alpha \sim {\mathcal N}^+(1, 2)$ (defined in Section \ref{sec:system}). We sample $5,000$ demand vectors from this distribution and present them as a scatter plot, where the heatmap shows sample density. The density of samples around location $(x, y)$ represents the probability that the request rate of $a$ and $b$ will be approximately equal to $x$ and $y$, respectively.

Intuitively, it is more important to ensure the system's service of the most frequently visited demand vectors. Similarly, the system's storage capacity should not be increased if the frequency of ``uncovered'' demand vectors is too low. For example, \autoref{fig:fig_coveredmass_examples} (c) depicts the capacity region of the replicated system $S(a,a,b,b)$. The size of this region is twice that of the region covered by the three-server systems (\autoref{fig:fig_coveredmass_examples} (a-b)), which means that this system can serve twice as many demand vectors. However, recall that both systems were designed to serve demand with a maximal request rate $\lambda=2\mu$. The coded system, $S(a,b,a+b)$, appears to provide the best coverage of this demand per its storage cost. This example demonstrates how the model would, ideally, be used to configure the capacity of a storage system and its redundancy scheme.

\begin{SCfigure}[2][t]
  \includegraphics[width=0.4\columnwidth]{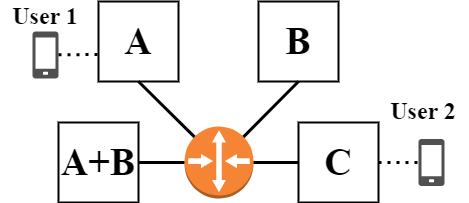}
  \caption{Schematic system with two users and four nodes.}
  \label{fig:service_types} \vspace{-5mm}
\end{SCfigure}

\section{Motivation}
\label{sec:motivation}

The model described in Section \ref{sec:model} represents a simplified system abstraction. In reality, the service capacity of the servers is limited not only by their storage and network bandwidth, but also by the size of their buffer queues. The system's users are dispersed geographically, in a way which affects their connectivity and might also affect their preference for specific objects. Even in a small-scale edge system, the number of objects and users might scale to many millions and thousands, respectively. 

Our goal in this work is to evaluate the applicability of the model to realistic systems. Specifically, we will identify the abstractions that that entail inherent gaps between the model and reality, provide means for bridging some of these gaps, and explain the challenges in addressing others.

\autoref{fig:service_types} depicts our basic system model. Each node is deployed in a static location, and the nodes are connected by a high-bandwidth network. Requests originate from \textit{user devices}, which are dispersed across the system and are connected to the nearest node by a low-bandwidth network. The user’s request enters the system via the nearest edge node, which serves as its \textit{access node}. For example, in the system in \autoref{fig:service_types}, $node_A$ and $node_C$ are the access nodes of users 1 and 2, respectively. 

If the access node stores the requested object, the request is served \textit{locally}: the access node sends the object directly to the user. Otherwise, the request is served \textit{remotely}: the access node forwards the request to the node storing the object, retrieves the object from the remote node, and forwards it to the user. A remote request thus waits in two queues to be served, instead of one. In \autoref{fig:service_types}, if both users request object $A$, then the object will be served locally to $user_1$ and remotely to $user_2$. If $user_1$ requests object $B$, it can be served remotely by $node_B$ or \textit{collaboratively}, from its recovery set on $node_A$ and $node_{A+B}$.  Next, we describe the actual systems used in our evaluation.

\section{Methodology}
\label{sec:system}




\textbf{Realistic edge system.}
We implemented (in C++) a small edge storage system that consists of two modules: the \textit{client} generates requests for specific objects, and the \textit{server} stores the objects and serves the requests. 
The clients and servers were deployed on physical programmable radio nodes of the ORBIT network research testbed~\cite{ORBIT}, which has been used extensively in the development of distributed systems and algorithms \cite{bronzino2019novn,gokalgandhi2019distributed,paris2011correlation}. Each node is equipped with 2 Intel Xeon CPUs, 192GB RAM, and a 256GB SSD, and the objects were stored in an in-memory Redis database. 

We used a cluster of seven neighboring nodes and deployed six servers, each on a distinct node, with six clients on the remaining node. Each client was assigned to one node as its access node, and the objects were distributed between all the servers. All nodes were connected to one another and communicated via TCP. 
To represent a widely distributed system, we limited the bandwidth between nodes to 1Gbps with a propagation delay of 1ms, and the delay between a user and its access node was 10ms. 
Requests were queued at each node and were handled by a pool of 100 threads. The queue in each node had a timeout of 100ms, and requests with a round-trip time greater than 200ms were discarded. As assumed for the model in Section \ref{sec:model}, equal node storage and capacity and object size are used. An object size of 64KB was employed, which is considered a medium size in the context of edge systems \cite{yang2022c2dn}.

\textbf{Simulated large-scale system.} 
To enable larger-scale experiments and more flexibility, 
we used EdgeCloudSim ~\cite{EdgeCloudSim} (an extension of CloudSim~\cite{CloudSim})---a multi-tier edge system simulator. This popular tool ~\cite{ecs_ref1,ecs_ref2,ecs_ref3,ecs_ref4,ecs_ref5} simulates geographically dispersed edge nodes and devices, the network bandwidth between them and the cloud backend, as well as queueing delays at each level of the hierarchy. We extended it to support object read requests (rather than compute jobs), implemented a mapping of objects to nodes, added a collaborative service for coded objects, and refined the implementation of the M/M/1 queues to be more realistic. 
Object sizes and network parameters were the same as in the ORBIT deployment. 

\textbf{Toy setting.} For our qualitative analysis, we recreated the toy systems $S(a,b,a+b)$ and $S(a,a,b,b)$ from Sec.~\ref{sec:model}. Our experiments consist of three and four nodes, respectively, with 2 users accessing each node. The users access two groups, $a$ and $b$, of three objects each: $a_1, a_2, a_3$ and $b_1, b_2, b_3$, respectively. Each node stores three objects according to the redundancy scheme. Namely, the three nodes in the coded system store group $a$, group $b$, and the parities $a_i+b_i$, respectively. In the replicated system, two nodes store group $a$ and two nodes store group $b$. The group request rates are $\lambda_a$ and $\lambda_b$ (normalized to $\mu_{n}$), with requests to each group distributed uniformly between its objects. We used a total of 90-250 traces and their corresponding demands in each workload (depending on the system and allocation) for the toy setting. 

\textbf{Realistic workload.}
Request distributions are known to exhibit a long ``tail'', i.e., some objects are very popular, and the majority of the objects are requested very rarely. Such distributions are commonly modeled by Zipf law~\cite{fan2023dynamic, EdgeKV, PopularityUserGeneratedContent:ChaKR09,YCSB}:  
the fraction of requests targeting the $i$th most popular object is proportional to $1/i^\alpha$, where $\alpha$ is the \textit{skew index} and $i$ is the object's \textit{rank}. The skew in content popularity increases with the index, $\alpha$. The portion of all requests that target a specific object depends on $\alpha$ and its rank $i$. 

Our workload consists of a set of individual request traces that follow the Zipf distribution, each trace representing a demand vector. Each request includes a timestamp, user ID, and the requested object. 
In each individual trace, the request rate generated by each user was determined by the Normal distribution $\mathcal{N}(0.8\mu,(0.2\mu)^2)$, where $\mu$ is equal for all nodes. 
In other words, the average $\lambda$ for the \emph{entire} workload is 80\% of the system's service capacity.
User requests arrive to each node 
as a Poisson process. 
In each trace, the popularity rank of the objects was determined by a random permutation of the objects. 
For the given skew, each request was mapped to an object which was sampled by its rank.

\textbf{User-inspired workload.}
We used the ``World of Warcraft (WoW)'' avatar-history dataset (WoWAH) ~\cite{WOWAH} to create a user-inspired workload. It contains user data collected 
in the servers of the popular online multiplayer game. The dataset includes basic information, in 10-minute granularity, about the user avatars and their approximate locations on the world’s map. We used this information to generate fine-grained locations and data objects representing map segments and neighboring avatars. Based on these, we generated a sequence of user requests for the objects relevant to each avatar's location. The original dataset consists of multiple \emph{logged periods}. We chose two of the longest contiguous periods, each reflecting a demand distribution for slightly less than a month. The details of these periods are given in Table~\ref{tab:wow}.

\begin{table*}
\centering
\begin{tabular}{|c|c|c|c|c|c|c|}
\hline
Period & Date & Duration & Included traces & Unique objects & Users & Mean request rate \\ \hline
1 & 6/2007 & 27 days & 1621 & 2913 & 1000 & 14326 reqs/sec\\ \hline
2 & 3/2008 & 21 days & 883 & 2875 & 1000 & 11611 reqs/sec\\ \hline
\end{tabular}
\caption{Traced periods used for the user-inspired workload.}
\label{tab:wow}
\end{table*}

We calculated the average request rate in each period, and included in our workload the traces of the 10-minute intervals whose request rate is between 0.8 and 1.2 of the average for each traced period. Since we treat a traced period as a demand distribution, we used the same object allocation for all the traces in each period. 
As a result, in each experiment (and corresponding demand vector), some objects were not requested. We scaled the nodes' capacities such that the average request rate is $0.8\mu$ i.e., 80\% of the cumulative service capacity of all the nodes in the system.

\textbf{Comparing with the model.} Our goal is to evaluate the model's ability to predict whether a given system will be able to serve a specific demand vector. 
We thus converted each individual trace into a demand vector by calculating $\lambda_i$ for each object $i$. We used the capacity-region model to classify this vector as inside or outside the service region. The model was implemented in Python based on the algorithm described in~\cite{CapacityRegion}. The implemented model also calculates the service cost and each node's \textit{load}---the percentage of its service capacity that is being utilized under the given demand.

To compare this classification to the results of the real and simulated systems, we defined a ``successful run'' as an experiment in which less than 2\% of the requests were discarded. This threshold\footnote{The results were similar with thresholds between 1\% and 3\%.} represents the system's tolerance to dropping some requests, and reduces the sensitivity of our analysis to arbitrarily bursty request combinations in the trace. We calculated the \emph{correlation} between the model and the systems for the entire workload. Specifically, we used Pearson's correlation coefficient for binary variables, which is also known as Phi coefficient or Matthew's Correlation Coefficient~\cite{powers2020evaluation}.  Typically, a score of $0.4$ or higher indicates a moderate correlation and a score of $0.7$ or higher indicates a strong correlation. Results are considered statistically significant if the p-value is lower than $0.05$. The results presented in this paper all have a p-value lower than $0.001$.

\section{Theory vs. Practice}
\label{sec:analysis}

\begin{figure*}
    \centering
\subfloat[Model (coding)]{
    \centering
    \setcounter{subfigure}{0} \includegraphics[width=0.5\linewidth,height=1.15in,keepaspectratio=true]{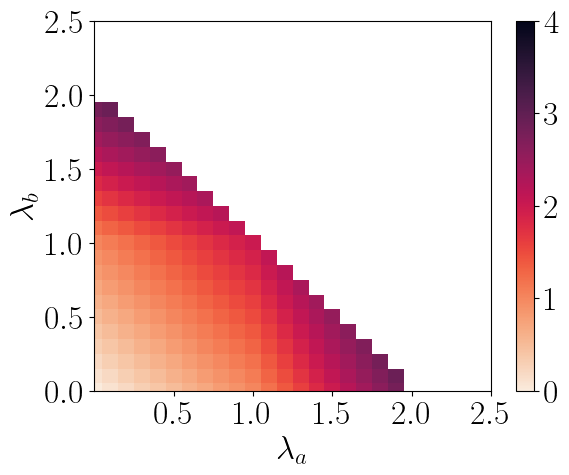}
}
\hspace{0.05\linewidth}
\subfloat[Simulated (coding)]{
    \centering
    \setcounter{subfigure}{2} \includegraphics[width=0.5\linewidth,height=1.15in,keepaspectratio=true]{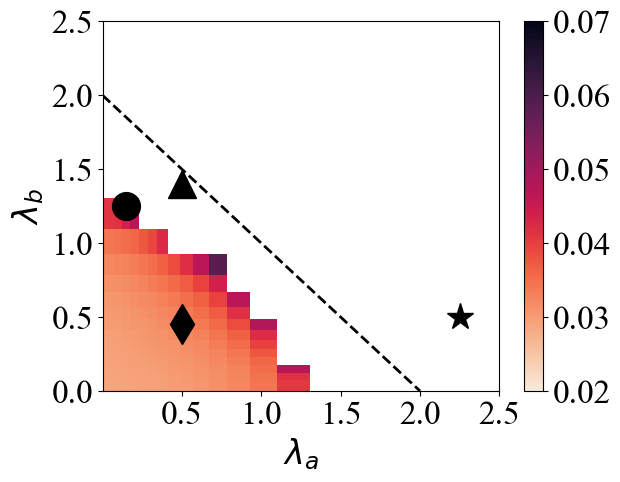}
}
\hspace{0.05\linewidth}
\subfloat[Real (coding)]{
    \centering
    \setcounter{subfigure}{4} \includegraphics[width=0.5\linewidth,height=1.15in,keepaspectratio=true]{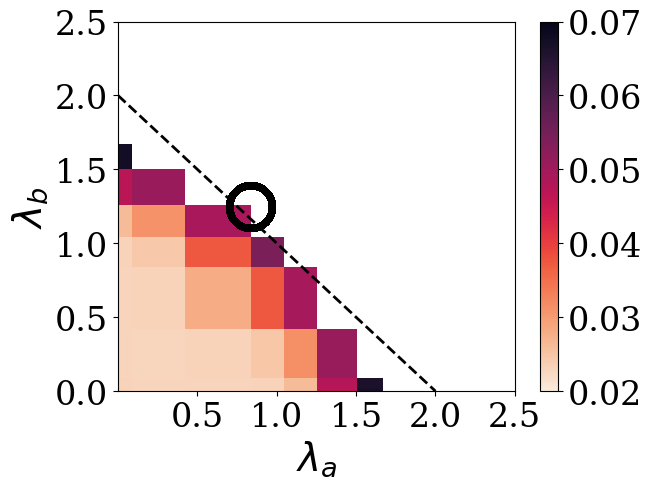}
}
\\
\subfloat[Model (replication)]{
    \centering
    \setcounter{subfigure}{1} \includegraphics[width=0.5\linewidth,height=1.15in,keepaspectratio=true]{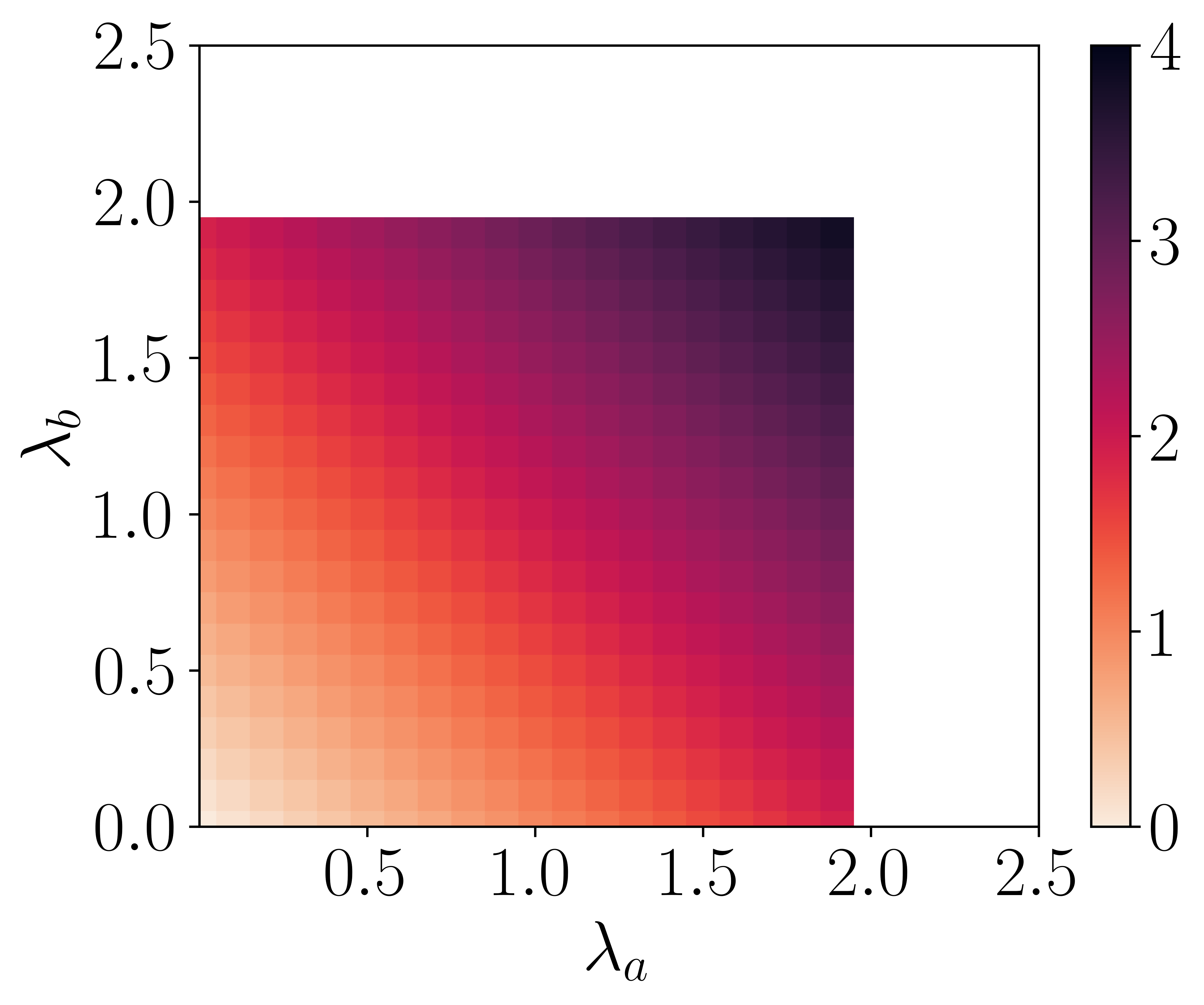}
}
\hspace{0.05\linewidth}
\subfloat[Simulated (replication)]{
    \centering
    \setcounter{subfigure}{3} \includegraphics[width=0.5\linewidth,height=1.15in,keepaspectratio=true]{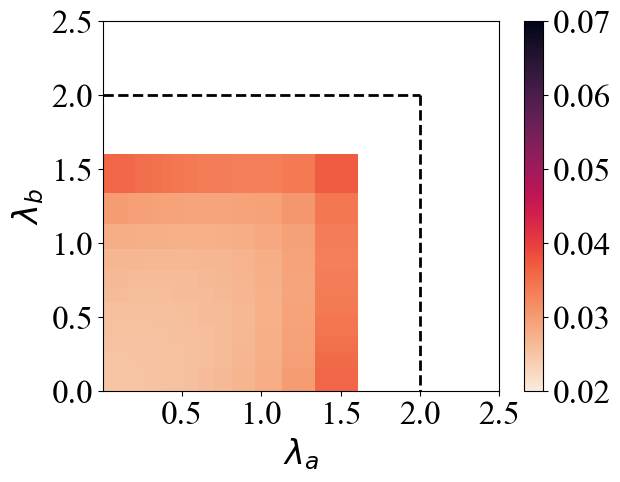}
}
\hspace{0.05\linewidth}
\subfloat[Real (replication)]{
    \centering
    \setcounter{subfigure}{5} \includegraphics[width=0.5\linewidth,height=1.15in,keepaspectratio=true]{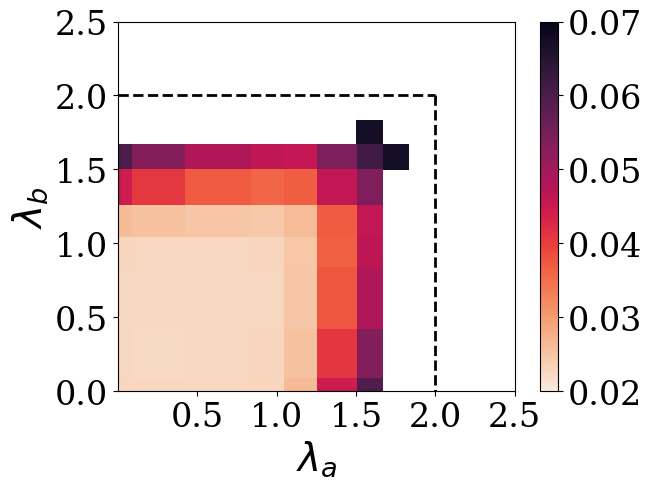}
}
    \caption{The service cost of the model and the latency of the simulator and real system, in a setup with coding and with replication. The dashed line depicts the boundary of the service region, and the points mark the cases discussed in Section~\ref{sec:analysis}.}
    \label{fig:service_cost}
    \vspace{-4mm}
\end{figure*}

\subsection{Qualitative comparison}
\label{sec:Qualitative}

We begin with a qualitative comparison between the model, the simulator, and the real system in the toy setting. The results are presented in \autoref{fig:service_cost}.
For the model (\autoref{fig:service_cost}(a-b)), we present the service cost within each configuration's capacity region. For the simulated (\autoref{fig:service_cost}(c-d)) and real (\autoref{fig:service_cost}(e-f)) systems, we present the average latency in the successful runs. Unsuccessful runs are represented by the white areas. 

The resemblance between the service region and the successful runs is evident, demonstrating the potential of the model to become a useful tool for predicting the service capacity of real systems. We also note increased latencies of the systems in the areas which were characterized by high service costs in the model. 
Indeed, a high service cost indicates that more objects were downloaded when serving the demand, increasing the load on the nodes and, respectively, the queueing and network delays in the system. 

The results of this experiment (also supported by our next experiment) show that the model's predictions tend to be optimistic, i.e., the service region covers some demands that the system could not serve. 

We first characterize the demands for which the model made \textbf{accurate predictions}. The first type is demands outside the service region that resulted in unsuccessful runs in the systems (e.g., $\bigstar$). This was typical of demands whose skew was too high or whose cumulative request rate in one of the nodes exceeded the node's service capacity $\mu$.
In general, due to the model's `optimism', its negative predictions (that a point is outside the service region) were highly accurate. 

The second type is demand vectors that imposed a low load on the system and were thus far from the service region's boundaries (e.g., $\blacklozenge$). In the real system and the simulator, the respective traces experienced low latencies thanks to the low traffic and short queueing delays. These demand vectors were ``easy'' to predict, and the model's positive predictions (that a point is inside the service region) were highly accurate.

The most common cause of the model's \textbf{inaccurate predictions} was when the model predicted that demand could be served, while the experiment resulted in an unsuccessful run (e.g., $\blacktriangle$). This misprediction resulted from two inherent gaps between the model and reality, each representing a major challenge in modeling edge-based systems.

\begin{figure}[t!]
    \centering
        \includegraphics[width=0.65\linewidth]{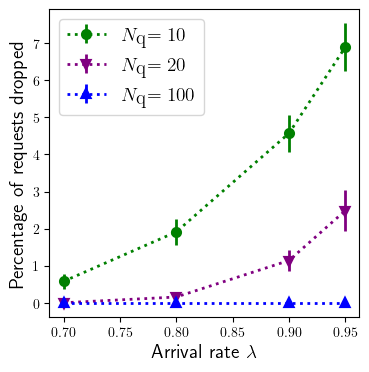}
    \vspace{-1mm}
\caption{The percentage of dropped requests at an $M/M/1$ queue with service rate of $1$. The markers and error bars depict the mean and standard deviation of 16 experiments with $10,000$ requests, respectively.}
    \label{fig:drop_rate_mm1}\vspace{-3mm}
\end{figure}

\underline{Gap 1: queue modeling.} Recall the model focuses only on identifying requests that cannot be served within a finite time frame. 
Real systems are limited by a \emph{concrete} finite queue size, and requests are dropped requests when the queue is full~\cite{sreekumar2020position,tang2022latency}. We demonstrate this in ~\autoref{fig:drop_rate_mm1}. It portrays  a single $M/M/1$ queue behavior with a service rate $\mu=1$. It highlights the percentage of dropped requests, varying with different buffer sizes $N_{\textrm{q}}$ and varying request arrival rates $\lambda < 1$.
To determine the dropped request percentage, we simulated the queue and examined the completion of 10,000 requests. For each combination of $N_{\textrm{q}}$ and $\lambda$, we repeated this experiment 16 times, plotting the average and standard deviation (represented by error bars).

The figure shows larger buffers reduced the percentage of dropped requests (the buffer of size 100 behaved like an infinite buffer in this setting). Meanwhile, the likelihood of a request being dropped increased as the request arrival rate approached the queue's service capacity ($\lambda \rightarrow 1$). As a result, we anticipate lower model accuracy for demand vectors near the boundaries of the system's capacity region.

In addition to finite queue size and in contrast to the model's assumptions, a request in a real system might wait in multiple queues, including those of the network interface. 
Generally, real systems use techniques such as prioritization and reservation techniques to  address congested queues~\cite{sreekumar2020position}.

\underline{Gap 2: access rate granularity.} 
The model assumes a steady request rate for each object. However, real traces may feature short yet intense bursts for specific objects, potentially causing some queues to overflow. These demands lie near the capacity-region boundaries, but the model can only identify the burst if the demand vector is computed for a sufficiently short trace. For example, consider two users that request an object for a total of 1000 times, at requests rates of 1 per msec and 1 per usec. If the demands are given in per-second granularity (i.e., 1000 reqs/sec), the model will not distinguish between them.  The model could effectively predict the system's ability to handle demands with differing cumulative request rates ($\lambda$) and object-ranking combinations. However, its predictions were less accurate when the popularity was extremely skewed (near region boundaries).

Demand vectors in the region's outskirts of successful runs (e.g., $\bullet$) deserve special attention as they impose high latencies, indicating a high load on nodes. Due to the model's optimism, it often accurately classified these points as within the capacity region. However, the system's ability to serve these demands is susceptible to sudden bursts that might result in unsuccessful runs.
For a similar reason, points served by the system under high load may have represented demands outside the service region, which the system managed to serve thanks to a favorable combination of concurrent requests (e.g., $\bigcirc$). Thus, the model's predictions for points near the boundary of the capacity region tended to be less accurate. 

\textbf{Improving the model's applicability.} Our observations suggest the existence of a \textit{gray region}---an area near the capacity region's boundary where the model's predictions are less accurate. 
We propose a simple method to identify this region to enhance the model's applicability. We noticed that demands in the gray region were those for which at least one of the system’s nodes became overloaded and dropped requests. Therefore, we define a node's \emph{load} as the portion of its capacity utilized to serve a demand, and the system’s \textit{max load} as the load of its most loaded node.

Max load represents the proximity of a demand vector to the capacity region's boundary, making it suitable for a parametric definition of the gray region. A gray region with a \textit{width} of $w$ includes demands with a max load $\geq 1-w$. For instance, a gray region width of 5\% includes all demands where the system’s max load is 95\% or higher. We later demonstrate that excluding demands within the gray region increases the model's accuracy.

\subsection{Quantitative analysis}
\label{sec:quantitative}

\begin{figure*}[t]
    \centering
%
    \begin{subfigure}[t]{0.85\linewidth}
        \centering
        \includegraphics[width=\linewidth]{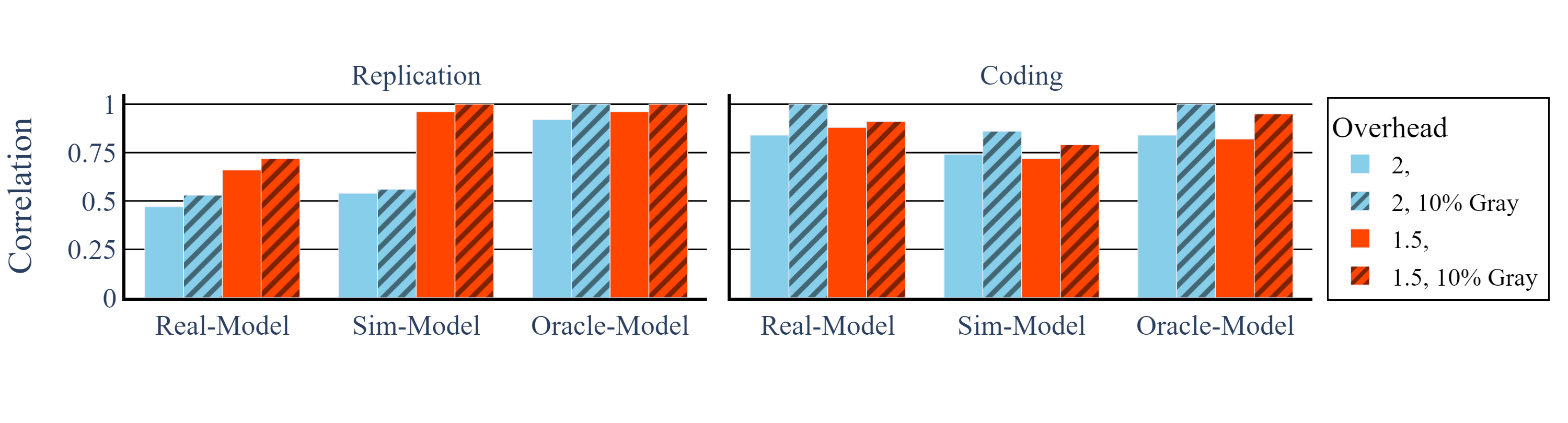}
            \vspace{-11mm}
        \caption{$\alpha=1$}
        \label{subfig:cor_z1}
    \end{subfigure}
    \begin{subfigure}[t]{0.85\linewidth}
        \centering
        \includegraphics[width=\linewidth]{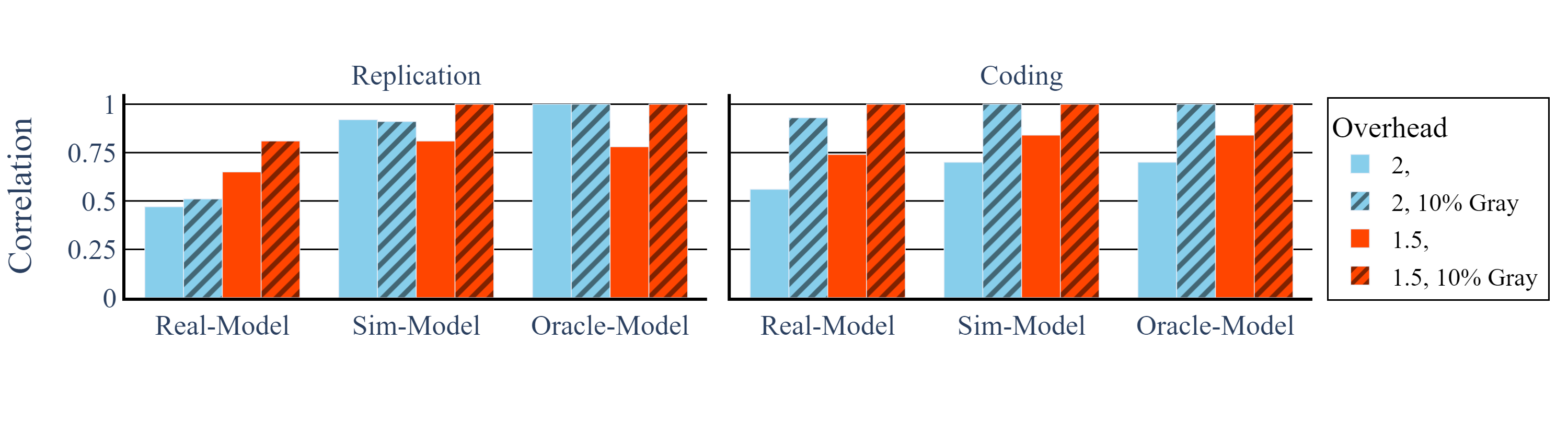}
            \vspace{-11mm}
        \caption{$\alpha=0.75$}
        \label{subfig:cor_z0.75}
    \end{subfigure}
    \caption{Correlation between the model and each of the two systems for each system setting for $\alpha=0.75,1$. `Oracle' represents ideal request routing in the simulator.}\vspace{-3mm}
    \label{fig:cor}
\end{figure*}

We observed a system comprising six nodes under a Zipf workload with $\alpha \in \{0.5, 0.75,1\}$, 60 objects, and two storage overhead values (1.5 and 2). For each setting, we generated 50 traces, each lasting 3 minutes, and corresponding demand vectors. 
\autoref{fig:cor} (a) illustrates the correlation between each system and the model in the different settings for $\alpha=1$.

The correlation was typically higher for coding than replication, especially in the real system. This is attributed to the enhanced recovery flexibility inherent in the coding scheme, in contrast to replication with an equivalent overhead. For example, with an overhead of 1.5, only half of the data objects had replicas,  but all data objects had respective recovery sets with which they could be served. This flexibility improved the model's accuracy as the system's behavior exhibited less sensitivity to bursts in demand for specific objects (Gap \#2).

To evaluate the benefit of identifying the gray region, we recalculated the correlation only for points identified by the model as being outside the gray region. The results for a gray width of 10\% are included in \autoref{fig:cor}. This refinement improved the correlation by up to 0.12 and 0.05 for coding and replication, respectively. At the same time, the portion of demand vectors excluded by the model was 15\% and 9\% for coding and replication, respectively. Both the correlation and the quantity of excluded vectors increased with the gray width. 

\autoref{fig:cor} (b) shows the results for $\alpha=0.75$. Due to the reduced skew, the correlation between the model and the systems is generally higher, especially with replication and low overhead. The trend is similar for $\alpha=0.5$, omitted due to lack of space.

\begin{figure}[t!]
    \centering
{
    \centering
\includegraphics[width=0.6\linewidth,valign=t]{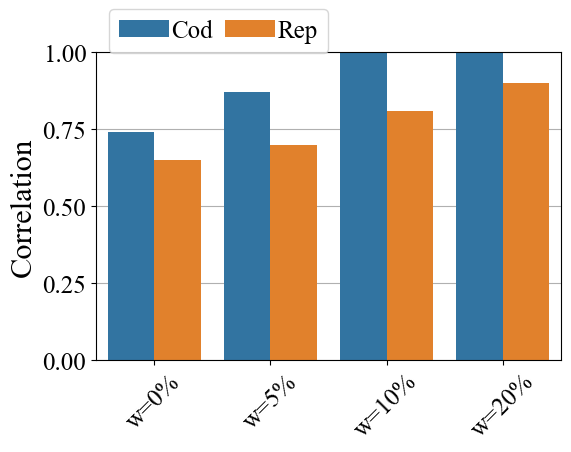}
}
    \caption{Real-Model correlation as a function of the gray area width, $w$, for an overhead of 1.5 and $\alpha=0.75$.}
   \label{fig:gray-cor}
    \vspace{-6mm}
\end{figure}

\begin{figure*}[t!]
    \centering
\subfloat[Coding]{
    \centering
\includegraphics[width=0.47\linewidth]{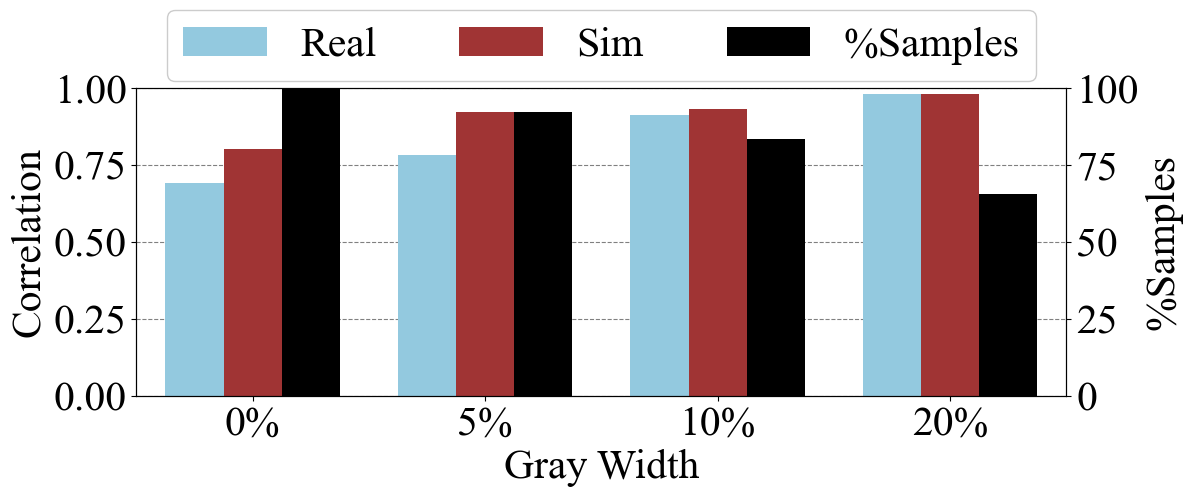}
}
~
\subfloat[Replication]{
    \centering
    \includegraphics[width=0.47\linewidth]{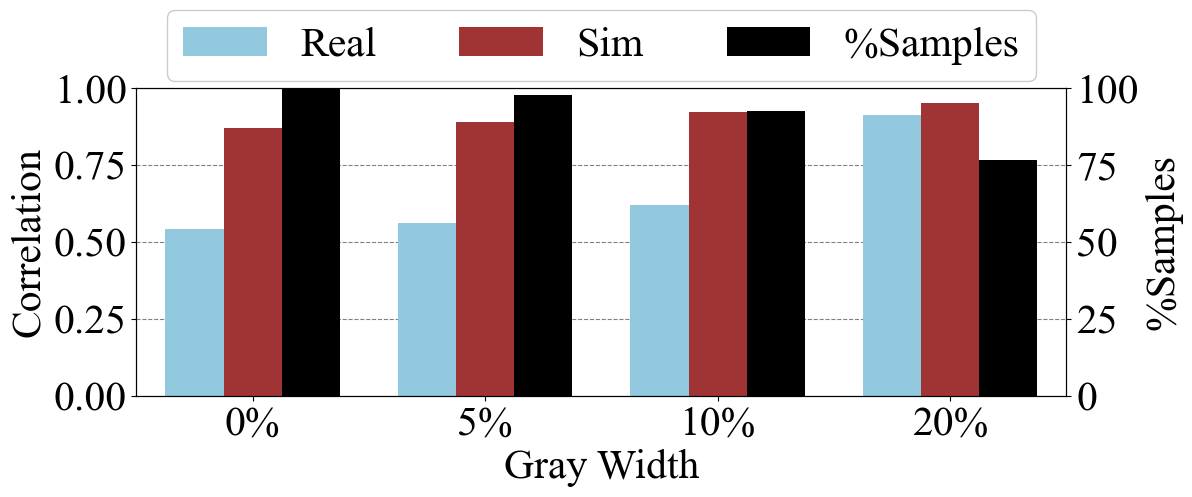}
}
    \caption{Correlation as a function of gray region width for all workload-system combinations.}
    \label{fig:gray_orbit_small}\vspace{-1mm}
\end{figure*}

Next, we examine the correlation between the model and the system as a function of gray region width $w$. \autoref{fig:gray-cor} shows the correlation of the realistic system to the model as the size $w$ increases for an overhead of 1.5 and skew $\alpha=0.75$. Again, the correlation was higher for coding than for replication due to the flexibility in the coding scheme.
As expected, the correlation increased as more points were removed from the vicinity of the boundary. 
However, this came on the expense of demand vectors excluded from the service region. 

We demonstrate this tradeoff in ~\autoref{fig:gray_orbit_small}. The figure shows the correlation between the model and the systems with different gray area widths, where each column represents the aggregate results for all (six) overhead-skew combinations. 
On the secondary Y axis, we show the portion of the demand vectors outside the gray area for each width, i.e., those included in the calculation.
The results show that when the samples in the gray area are excluded, the correlation between the model and the systems increases dramatically. This is expected, as most of the remaining points are of types shown to be predicted correctly, i.e., $\blacklozenge$ and $\bullet$.

Mispredictions in which the system successfully served demands outside the service region (e.g., $\bigcirc$) were scarce. In this experiment, they accounted for only 9.4\% of all the mispredictions in the realistic system (including those in the gray region). Specifically, such cases occurred mostly for overhead of 1.5 and $\alpha=1$, i.e., highly skewed object popularities and relatively low redundancy. This type of misprediction is the most difficult to eliminate, as it strongly depends on fine-grained system configuration parameters. To verify the robustness or our results, we repeated the experiment with overhead 1.5 and $\alpha=0.75$ with a workload of 400 samples instead of 50. The results 
were the same as in \autoref{fig:gray_orbit_small}.


\underline{Gap 3: real-time system state.}  Ideally, when requests are served remotely, each request will be served by the least loaded node at the time of its arrival. In reality, distributed systems \cite{cassandra,hdfs} (as well as our own implementations) use periodic messages exchanged between nodes to estimate one another's current state.  This approach may lead to requests being forwarded to a node with a full queue, resulting in the request being dropped, even though a less loaded node could have processed it.

To demonstrate this gap, we repeated the simulator runs in an \emph{Oracle} mode. In this mode, each node was aware of its peers' current queue size when making routing decisions. The result for $\alpha=1$ is shown in \autoref{fig:cor} (a). 
The correlation between the simulated oracle and the model was 13\% and 25\% higher than the correlation between the simulator and the model, for coding and replication, respectively. Smaller differences were observed for lower $\alpha$ values of 0.75 (\autoref{fig:cor} (b)) and 0.5 (omitted), further demonstrating the effect of highly skewed demand discussed in Gap \#2.
When calculated for all workload-system combinations, the simulated oracle showed 5\% and 8\% higher correlation, for coding and replication, respectively. 

Bridging the real-time state gap requires a combination of modeling and system-level efforts. The model might add a probability for routing error, where it will ``intentionally'' assign a configurable number requests to loaded nodes. System-level solutions to bridge this gap are related to existing efforts for load balancing~\cite{
khan2015different}, monitoring~\cite{ali2019probabilistic}, and predictive routing based on machine learning~\cite{liu2021load}. 
Task-redundancy is often used to compensate for sub-optimal routing.

\underline{Gap 4: Geo-location and connectivity.}
Recall that the model does not distinguish between local and remote requests, thus neglecting the effect of request routing. To show this, we designed an experiment with extreme dependencies between user locations and their preference for objects.
We used the same distribution with three user behaviors: \textit{baseline}, where users requested objects regardless of their location; \textit{local}, where users requested only objects stored on their access node; and \textit{remote}, which was equivalent to the local behavior, but with each user shifted to an adjacent access node, making all requests remote. We generated 300 traces of each assignment type and used the simulator to run them with overhead 1.5 in replicated and coded systems.

\begin{table}
\center
\begin{tabular}{ |c|c|c|c|}
\hline
 \textbf{User behavior} & Baseline & Local &  Remote   \\ 
 \hline
 \textbf{Correlation} & 0.81 & 0.84 & 0.94 \\ 
 \hline
   \textbf{Correlation (10\% Gray)} & 1 & 0.8 & 0.96 \\ 
 \hline
\end{tabular}
\caption{Correlation for the geolocality experiment.
}\label{tab:geo}
\vspace{-4mm}
\end{table}

The results are summarized in ~\autoref{tab:geo}. The correlation for remote behavior was 0.94, which is 0.13 higher than the baseline, and 0.1 higher than the local behavior. When all requests were remote, the system distributed them between the nodes storing their objects and those storing their replicas or recovery sets, according to the load on each node. This essentially balanced the load evenly between all nodes, similar to the model. When a 10\% gray region was applied, the baseline correlation reached 1, similar to the remote correlation of 0.96. The correlation for the local behavior was less affected by the gray region.

To bridge Gap \#4, a layer of routers could be added to the storage nodes. Routers would have finite capacity to stream data objects and can possibly limit the capacity region. We leave such extension to future work.

\begin{figure*}
  \centering
 {
    \centering
\includegraphics[width=\linewidth]{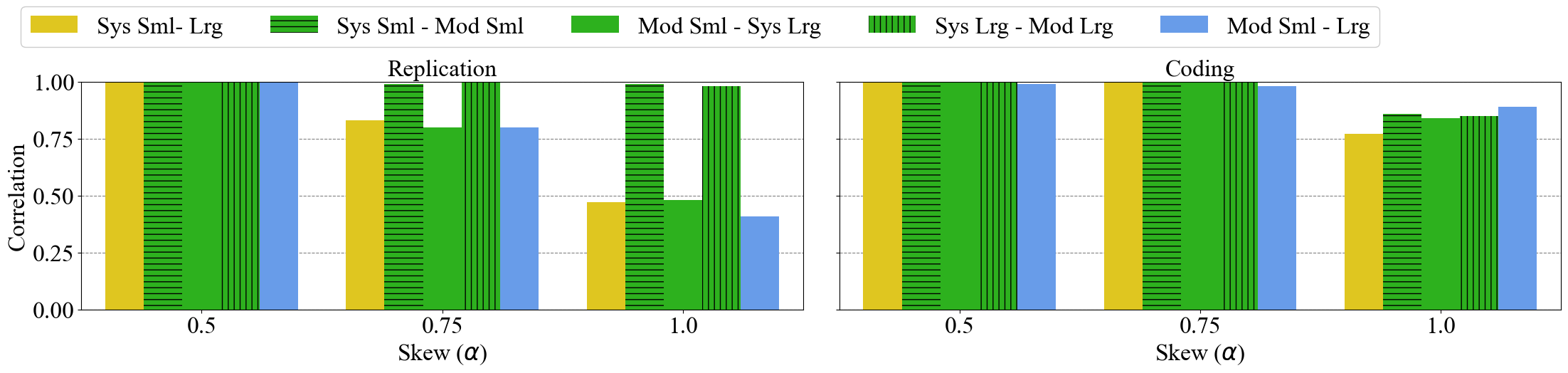}
}
\vspace{-4mm}
\caption{Correlation between systems with 10 nodes and 100 objects (Sml) or 1000 objects (Lrg). $w=10\%$.}
    \label{fig:sml2large}
\end{figure*}

\underline{Gap 5: model and system size.}
Modeling large-scale systems can be challenging due to their complexity and scale, and different models might be needed for different system sizes~\cite{shmuel2018performance, ScaleAndPerformance:HowardKM88, PerformanceModelingToDesignLargeScaleSystems:BarkerDH09, WorkloadMmodelingForComputerSystemsPerformanceEvaluation:Feitelson15}, creating modeling gaps.
Recall the size of the demand vector equals the number of objects, $k$. To determine whether a system can serve the expected demand distribution, the model must evaluate a collection of such vectors, indicating whether they are within or outside the system's service region. For example, consider a system with 130 million unique objects~\cite{yang2022c2dn}. Querying the model for 50 demand vectors would produce 6.5 billion request rates. A natural question is whether we can predict the behavior of a large system by using the model on a small system that represents it.

To answer this question, we compared two sets of Zipf workloads, with 100 (small) and 1000 (large) objects, in a simulated system with 10 nodes and a storage overhead of 1.5. 
We first generated the traces and object allocation for the large system. We then derived the traces and allocation for the small system, as follows. In each node in the large system, we aggregated groups of 10 data objects from those stored in this node and represented each group as a single (aggregated) data object in the small system. We then generated the respective replicas or recovery sets in the small system. The trace in the small system contained the same requests as those in the large system, after converting the original objects to the aggregated ones. We obtained the results of the model and the simulator for both large and small systems, and calculated the correlation between the results of all experiment combinations. 

\autoref{fig:sml2large} shows the correlation in these experiments, with a gray area width of $w=10\%$. We are most interested in the correlation between the model of the small system and the simulator of the large system (``Mod Sml - Sys Lrg''). The other columns serve as references. Interestingly, this correlation is high when using coding, and then using replication with moderate skew. However, when the system uses replication and the skew is high ($\alpha=1$), the correlation drops to 0.49. The reason is similar to that in Gap \#2: the aggregation of objects ``hides'' the demand peaks of the most popular objects, causing the model to misclassify the vector with respect to the capacity region. To see that this is the reason, note that the correlation between the model and the system of the same size (``Sys Sml – Mod Sml'' and ``Sys Lrg – Mod Lrg'') is always high. On the other hand, the correlation between the predictions on different sizes size (``Sys Sml – Lrg'' and ``Mod Sml - Lrg'') decreases with the skew. This gap might be addressed by a better representation of the large system, e.g., selective aggregation or sampling.


\begin{figure}
    \centering{
            \vspace{-8mm}        \includegraphics[width=0.7\linewidth, valign=b]{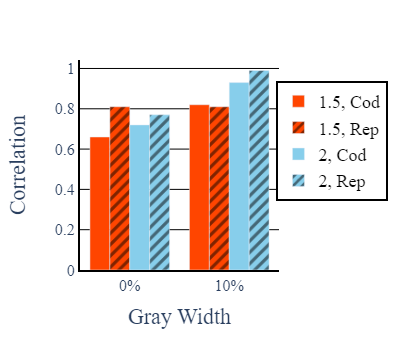} 
        \label{subfig:wow-cor}
}
    \vspace{-3mm}\caption{User-inspired workload in a system with 100 nodes, overheads of 1.5 and 2, and gray widths $w=0\%,10\%$.}
                \vspace{-4mm}
        \label{fig:wow}
\end{figure}
\textbf{User-inspired Workload.} We compare the model to the system with the user-inspired workload described in Section~\ref{sec:system}. We experimented on 1000 different traces from two periods. This workload is larger (up to 2913 objects) and more skewed: 80\% of the requests are to the 16\% most popular objects, compared to 40\% most popular objects for Zipf workload with $\alpha=1$. We simulated a system with 100 nodes and overheads of 1.5 and 2, and scaled the system's cumulative service capacity so that the average request rate is 80\% of the system's capacity ($\lambda=0.8\mu$). The results are for aggregating traces from both traced periods (see \autoref{tab:wow}).

\autoref{fig:wow} depicts the correlation between the model and the system with $w=0\%,10\%$. 
As in the Zipf traces, the correlation between the model and the system is higher when the storage overhead is higher, and improves with the addition of the gray region. This is to be expected, as this workload contains demand spikes for specific objects, spikes driven by user behavior (Gap \#2). We confirmed that the accurate request routing in Oracle mode increased the correlation by up to 0.12. Overall, despite the bursty request rate and high skew, the user-inspired workloads exemplified the same gaps we identified using the Zipf workload. 



\section{Related Work}
Despite being an indispensable component of modern software, there is limited literature on the analytical modeling of modern distributed storage systems. Most existing works, such as \cite{MethodOfLayers:RoliaS95, QueueingNetworkModelForADistributedDatabase:JenqKT88, ModelingOfDistributedDatabases:NicolaJ00, DatabasePerformanceEvaluation:OsmanK12}, tend to represent distributed storage systems as queueing networks. 
These models primarily aim to predict system performance through simulations or numerical calculations, capturing system behavior across a wide array of performance metrics. 
The mathematical analysis of such models is complex due to interactions between different queues, which raises issues such as concurrency and inter-queue dependency.
This is why, these models typically don't allow for deriving analytical expressions or insight on the system's performance.
An exception is the Fork-Join 
queueing models, employed for accessing distributed data \cite{FJModel:Varki01, DelayStorageTradeoff:JoshiLS14}. 
Their state space is typically smaller and more tractable than models based on queueing networks, but their applicability is limited to systems with specific redundancy and data access schemes.
To the best of our knowledge, the capacity-region model \cite{CapacityRegion} is the only one that provides a mathematical analysis of the service capacity of distributed storage systems.

\section{Conclusions and Open Challenges}

We compared the best available analytical model for edge-based storage service to real and simulated edge systems. Our evaluation revealed inherent gaps between theory and practice and showed how some of them can be partially addressed by a simple refinement of the model. Further refinements are required for addressing these gaps systematically. We expect to observe additional gaps when evaluating more sophisticated user behavior (such as \textsc{write} requests or caching of highly popular objects) and redundancy schemes. Analytically modeling  these effects remains an open challenge.
\section*{Acknowledgment}

We thank Harel Karni for his help in the extension of EdgeCloudSim. This research was supported in part by US-Israel BSF grant 2021613 and by ISF grant 807/20, NSF-BSF grant FET-2120262, and NSF grant CIF-2122400.

\bibliographystyle{abbrv}
\bibliography{bib,EdgeWorkloads,ErasureCodes,fogComputing,nsf,ref}

\end{document}